\documentclass[preprint,aps,12pt,preprintnumbers,eqsecnum,nofootinbib,superscriptaddress]{revtex4}
\usepackage{amsmath}
\usepackage{amsfonts}
\usepackage{amssymb}
\usepackage{amsthm}
\usepackage{mathtools}
\usepackage{physics}
\usepackage{hyperref}
\usepackage{xcolor}
\usepackage{graphicx}
\newcommand\bea{\begin{eqnarray}}
\newcommand\eea{\end{eqnarray}}
\newcommand\Pl{\rm{Pl}}

\allowdisplaybreaks

\usepackage{soul} 

\begin{document}

\title{A model of composite gravity with Pauli-Villars regulators}
	\author{Chris Li}
	\email{chrisli@berkeley.edu}
\affiliation{
Department of Physics, University of Virginia, Charlottesville, Virginia 22904, USA
}
\affiliation{
Department of Mathematics, University of California, Berkeley, Berkeley, California 94720, USA
}
\affiliation{
Department of Physics, University of California, Berkeley, Berkeley, California 94720, USA
}

\author{Diana Vaman}
\email{dvaman@virginia.edu}
\affiliation{
Department of Physics, University of Virginia, Charlottesville, Virginia 22904, USA
}
	\date{\today}

\begin{abstract}{We revisit a model of composite gravity, in the form of a reparametrization invariant, non-polynomial,  metric-independent action for scalar fields.  Previously, the emergence of a composite massless spin 2 particle, the graviton, was demonstrated by analyzing  a two-into-two scalar scattering  amplitude. Working in the limit of a large number of physical scalars and using dimensional regularization,  it was shown that the scattering amplitude had a pole corresponding to a graviton exchange,  provided that a certain fine-tuning was implemented; the Planck mass was determined as a function of the dimensional regularization parameter and a mass scale.  Here we demonstrate that the presence of the composite graviton is a robust feature of this model and not an artefact of the choice of regulator,  by replacing dimensional regularization with Pauli-Villars fields. The presence of the massless graviton is conditioned by a similar fine-tuning as before. This is arguably a more physical regularization, since the Planck mass now depends on the specifics of the Pauli-Villars regulator fields, e.g. their mass as well as their multiplicity.}
\end{abstract}
	
	\maketitle

\section{Introduction}

In this paper we are studying a model of  composite gravity in a scalar theory considered earlier by Carone, Erlich and Vaman \cite{Carone:2016tup,Carone:2017mdw}.\footnote{A similar model of composite gravity in a fermionic theory was discussed in \cite{Carone:2018ynf}, and another extension was given in \cite{Batz:2020swk}.} For an overview of this series of work and further insights see \cite{Carone:2019xot}. For the sake of completeness we will recall the key features of the model. The action for the theory resembles the  Dirac-Born-Infeld action with a vanishing gauge field, modulated by a potential function $V(\phi^a)$:
\begin{equation}
S=\int d^Dx\ \left(\frac{\tfrac D2-1}{V(\phi^a)} \right)^{\frac{D}{2}-1}
\sqrt {\bigg|\det \left(\sum_{a=1}^N \partial_\mu\phi^a \,\partial_\nu\phi^a 
+\sum_{I,J=0}^{D-1}\partial_\mu X^I \,\partial_\nu X^J\, \eta_{IJ}\right)\bigg|}.
\label{eq:S}\end{equation}
This action is reparametrization invariant. This  symmetry is gauge-fixed by identifying  the clock-and-ruler  fields $X^I$ with the corresponding spacetime coordinates
\begin{equation}
X^I=  x^\mu\delta_\mu^I \ \sqrt{\frac{V_0}{\tfrac{D}{2}-1}-c_1} , \ \ I=0,\dots,D-1 \, , \label{eq:staticgauge}
\end{equation}
and where $c_1$ is a counterterm  chosen to normal-order every occurence of $\sum_a \partial_\mu \phi^a \partial_\nu\phi^a$ in (\ref{eq:Sexpansion}) ({\em i.e.}, any loop which can be constructed by contracting the two $\phi^a$'s in $\sum_a \partial_\mu \phi^a \partial_\nu\phi^a$ is rendered zero by adding the 
counterterm $-c_1 \eta_{\mu\nu} $).
In order to analyze the theory perturbatively,  one writes $V(\phi)=V_0+\Delta V(\phi^a)$ and expands the action in (\ref{eq:S}) in powers of $1/V_0$. Another (simplifying) assumption is that $N$, the number of fields $\phi^a$ in the theory, is large, which justifies keeping only the leading terms in a $1/N$ expansion.  In the two-into-two scattering calculation of Ref.~\cite{Carone:2016tup}, this made the desired
diagrammatic resummation possible. 

The gauge-fixed action, expanded to second order in $1/V_0$, reads:
\begin{eqnarray}
S=\int d^Dx&&\left\{\frac{V_0}{D/2-1}+\frac{1}{2}:\sum_{a=1}^N \partial_\mu \phi^a \partial^\mu \phi^a: -\Delta V(\phi^a )\right. \nonumber \\ &&
-\frac{\tfrac D2-1}{4V_0}\left[
:\sum_{a=1}^N\partial_\mu\phi^a \partial_\nu\phi^a :
\,:\sum_{b=1}^N\partial^\mu\phi^b\partial^\nu \phi^b:
-\frac{1}{2}\left(:\sum_{a=1}^N \partial_\mu \phi^a \partial^\mu \phi^a : \right)^{\!\!2\,}\right] \nonumber \\
&& \left.-\frac{\tfrac{D}{2}-1}{2}\frac{\Delta V(\phi^a)}{V_0}:\sum_{a=1}^N \partial_\mu \phi^a\partial^\mu \phi^a:+\frac{D}{4}\frac{(\Delta V(\phi^a))^2}{V_0}+{\cal O}\left(\frac{1}{V_0^2}\right)\right\}.   \label{eq:Sexpansion}
\end{eqnarray}
For simplicity the potential was chosen to be  O$(N)$-symmetric and quadratic
\begin{equation}
\Delta V(\phi^a)=\sum_{a=1}^N\frac{m^2}{2}\phi^a\phi^a-c_2=\,:\sum_{a=1}^N\frac{m^2}{2}\phi^a\phi^a:\,\label{c_2}.  
\end{equation}
 Similarly to $c_1$, the role of the counterterm $c_2$ introduced in (\ref{c_2}) is to normal-order every occurrence of $\sum_a \phi^a \phi^a$.

\begin{figure}[h!]
\includegraphics[scale=0.5]{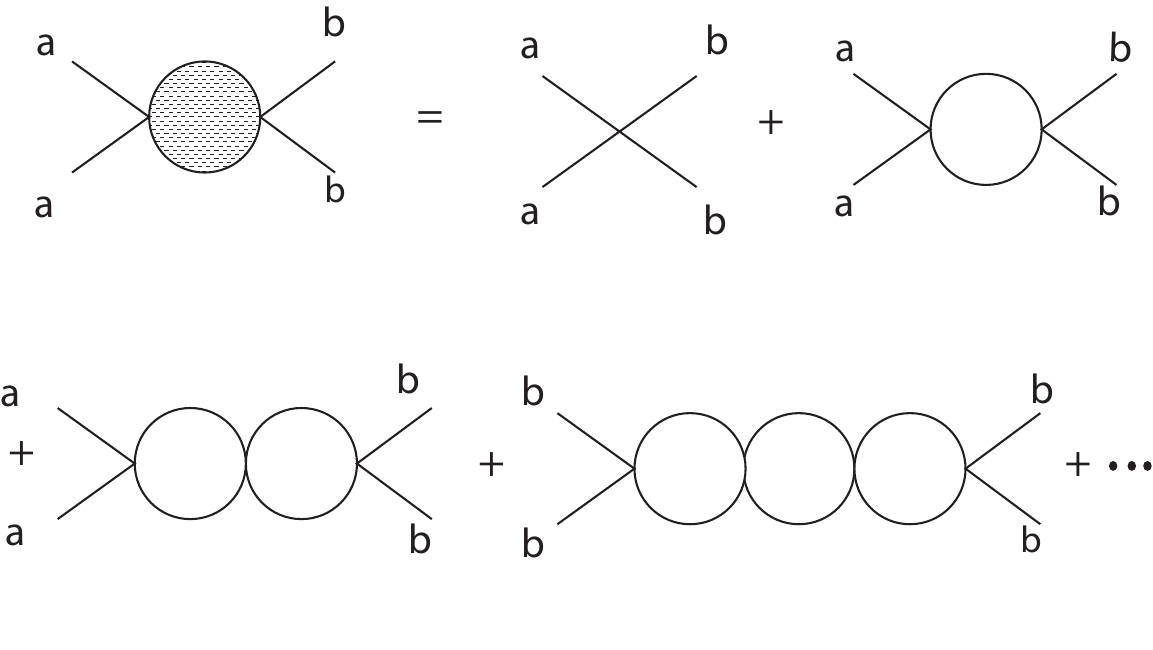} \caption{Two-into-two scalar scattering}\label{loops}
\end{figure}

The graviton pole was identified in \cite{Carone:2016tup} by considering the two-to-two scattering of $(\phi^a\, \phi^a\rightarrow \phi^b\,\phi^b) $ scalars in the large-$N$ limit\footnote{This treatment is similar to that of Suzuki~\cite{Suzuki:2010yp} who identified a composite gauge boson in a particular scattering process in a theory of emergent electromagnetism.} as shown in Fig.~\ref{loops}.

There are no other Feynman diagrams to consider (e.g. with  loops added to propagators and vertices) because of the counterterms $c_1$ and $c_2$. For more details, see Appendix A.

In \cite{Carone:2016tup}, dimensional regularization was used as a regulator of the loop integrals. The existence of a massless spin-two state (the graviton) being exchanged in this process
required the fine-tuned choice
\bea
V_0=-\frac{N(D/2-1)}{2} \frac{\Gamma(-D/2)}{(4\pi)^{D/2}} (m^2)^{D/2}    \,\, , \label{choice}
\eea
leading to the following expression for the scattering amplitude:
\begin{equation}
A^{\mu\nu|\rho\sigma}(q) =  -\frac{3 \,m^2}{D\, V_0}\, \left[(\tfrac D2-1) \,( \eta^{\nu\rho} \eta^{\mu\sigma} + \eta^{\nu\sigma} \eta^{\mu\rho}) 
- \eta^{\mu\nu} \eta^{\rho\sigma} \right] \, \frac{1}{q^2} +\cdots \,\,\, 
\label{eq:ampsol}
\end{equation}

Comparison with the corresponding graviton-mediated scattering amplitude in a free scalar theory
\begin{equation}
A^{\mu\nu|\rho\sigma}(q) = -\frac{M_{\Pl}^{2-D}}{D-2} \, \left[(\tfrac D2-1) \,( \eta^{\nu\rho} \eta^{\mu\sigma} + \eta^{\nu\sigma} \eta^{\mu\rho}) 
- \eta^{\mu\nu} \eta^{\rho\sigma} \right] \, \frac{1}{q^2} \,\,\, ,\label{mpl0}
\end{equation}
where $M_{\Pl}$ is the $D$-dimensional Planck mass, 
leads to the following identification 
\begin{equation}
M_{\Pl}= m\,\bigg[\frac{N \, \Gamma(1-\frac{D}{2})}{6\, (4 \pi)^{D/2} }\bigg]^{1/(D-2)} \,\,\, . \label{eq:mpscat}
\end{equation}
With $D=4-\epsilon$, requiring that the Planck mass be positive implies that the regulator $\epsilon$  is small and negative. The dimensionful constant $V_0$, as identified in (\ref{choice}), is 
however positive.  Lastly, with $V_0$ fine-tuned to the value in \eqref{choice},  $c_1- V_0/(\tfrac D2-1)=0$. 
This renders the clock-and-ruler fields, which were gauge fixed according to  \eqref{eq:staticgauge}, zero.  However, we note that we could have formally expanded the square-root determinant in the action \eqref{eq:S} even without the clock and ruler fields, using 
\bea
\sqrt{\det h_{\mu\nu}}=\sqrt{\det (h_{\mu\nu}-\eta_{\mu\nu}+\eta_{\mu\nu})}=\exp\bigg[\frac 12 {\rm{Tr}}
\bigg(\sum_{n=1}^\infty
\frac{(-1)^{n+1}}
n (h_{\mu\nu}-\eta_{\mu\nu})^n\bigg)\bigg].
\eea

Past related works include Sakharov's induced gravity \cite{Sakharov:1967pk},  where gravitational dynamics arises  by integrating out quantum fields coupled to a background metric,  resulting in an effective action which contains the EInstein-Hilbert term for the background metric.  In contrast,  here the starting point is non-metric, and the emergent gravity is not semi-classical, but fully quantum.  Other models of composite gravity  \cite{Ohanian:1969xhl} did not go beyond linearized gravitational interactions.  However the action proposed by \cite{Ohanian:1969xhl} bears a striking similarity with ours, when truncated to quartic order in fiels: in  \cite{Ohanian:1969xhl} the quartic interactions are of the form $t_{\mu\nu}t^{\mu\nu} +\# (t_{\mu}^\mu)^2$, with $t_{\mu\nu}$ the energy-momentum tensor of the non-gravitational field theory and $\#$ an adjustable parameter to ensure the existence of composite graviton.  The exchange of the "Goldstone boson" composite graviton was demonstrated in scattering processes by a similar resummation of loop diagrams as described earlier. However, the model proposed by  \cite{Carone:2016tup} goes beyond the quartic interaction terms.  Indeed, in\cite{Carone:2017mdw}  it was shown that by expanding the action \eqref{eq:S} to next order in perturbation theory (to order $(\phi)^6$),  the cubic graviton self interactions are reproduced as well.  For other related works see \cite{Terazawa:1976xx}.

The purpose of the current work is to investigate what would change if a different regulator is being used. Concretely, is the presence of a graviton pole dependent on the choice of regulator? This is a valid question given that our model \eqref{eq:S} is a higher-derivative (i.e. non-renormalizable) theory and physical scales such as the Planck mass depend on the regulator. Arguably,  it would also be preferable not to have the Planck mass depend on the dimensional regularization parameter $\epsilon$ as in \eqref{mpl0}, and instead have a more physical choice of regulator.

One thing is certain: in order to be able to find a composite graviton in the theory, we want to choose a regulator that will not break the general covariance of the model. For this reason we will introduce Pauli-Villars fields.

\section{Regularizing with Pauli-Villars fields}

We consider Pauli-Villars  fields $\Phi_{PV,i}^a$ which are bosonic and couple to background gravity in the same way as the physical scalars $\phi^a$. That means that, by eliminating the non-dynamical background metric  as in \cite{Carone:2016tup} via its equation of motion, we arrive at the same type of non-metric, non-polynomial action as encountered earlier in  
\eqref{eq:S}.  Specifically, the new action is
\bea
S&=&\int d^4x\ \frac{1}{V(\phi^a,\Phi_{PV,i}^a)}
\sqrt {|\det G_{\mu\nu}|},\\
G_{\mu\nu}&=&\sum_{a=1}^N \partial_\mu\phi^a \,\partial_\nu\phi^a +
\sum_{i=1}^{N_{PV}}\sum_{a=1}^N \partial_\mu\Phi_{PV,i}^a \,\partial_\nu\Phi_{PV,i}^a 
+\sum_{I,J=0}^{D-1}\partial_\mu X^I \,\partial_\nu X^J\, \eta_{IJ},\\
V&=&V_0+\frac 12:\sum_{a=1}^N\bigg(m^2 \phi^a \phi^a + \sum_{i=1}^{N_{PV}} M^2_i \Phi_{PV,i}^a\Phi_{PV,i}^a\bigg):.
\eea
The Pauli-Villars couple to each other and to the physical scalars in the same way that the physical scalars  coupled to each other.  We gauge-fix the clock-and-ruler fields as in \eqref{eq:staticgauge}, and proceed with the expansion of the action as in \eqref{eq:Sexpansion}.

The Pauli-Villars fields are the regulators of the theory,  and are defined with an unsual path integral functional determinant, as done by Diaz, Troost, van Nieuwenhuizen and Van Proyen \cite{Diaz:1989nx} (see also \cite{Anselmi:1991wb, Anselmi:1993cu})
\bea
\int \mathcal D' \Phi_{PV} e^{-\int \Phi_{PV}^T \cdot D_{(2)} \cdot \Phi_{PV}}=(\det D_{(2)})^{-\alpha/2} \,.\label{PV0}
\eea
Despite the resemblance with a Gaussian functional integral, the functional determinant is defined with an exponent which is not the usual -1/2. The path integral measure over the Pauli-Villars fields, which is responsible for this definition, is not invariant under shifts of the integration variables, so introducing sources  for the Pauli-Villars fields would lead to contradictions \cite{Anselmi:1991wb}. However, the sources are not necessary,  and we shall not introduce them. The exponent $\alpha$ in \eqref{PV0} is called the statistical weight of the Pauli-Villars field.
What this means for practical purposes is that a Pauli-Villars loop in a Feynman diagram will contribute to the scattering amplitude with a factor of $\alpha$ (relative to a similar diagram with physical scalar fields running in the loop).\footnote{This is similar to how Feynman diagrams with fermions loops are accompanied by a negative sign.
In more generality,  assume that the differential operator $D_{(2)}$ depends on background fields, and exponentiate to turn the expression in \eqref{PV0} into an effective action for the background fields.  The weight $\alpha$ becomes an overall factor multiplying the perturbative expansion of $\Tr\ln D_{(2)}$ in terms of the background fields.}
 {The reader may ask whether the weights of the Pauli-Villars fields are predetermined. The answer is: not necessarily, though there are natural choices for these weights that regularize the loop integrals,  as we shall see in the next couple of sections. The natural weights for our theory are sensitive to $N_{PV}$, the total number of Pauli-Villars fields present.}

More concretely, the evaluation of the scattering amplitude performed in Ref.~\cite{Carone:2016tup} to leading order in $N$ was done by summing up the Feynman diagrams in Figure \ref{loops}.\footnote{Note that as a result of including the counterterms $c_1$ and $c_2$ we do not have additional loops attached to the propagators or vertices that arise from \eqref{eq:Sexpansion}.}
To this end, define the external line factors
\bea
E^{\mu\nu}(p_1,p_2)\equiv-(p_1^\mu p_2^\nu+p_1^\nu p_2^\mu)+\eta^{\mu\nu}(p_1\cdot p_2+m^2),
\eea
and write the scattering amplitude as
\bea
i{\cal M}(p_1,a;p_2,a\rightarrow p_3,b;p_4,b)\equiv E_{\mu\nu}(p_1,p_2) \, i A^{\mu\nu|\rho\sigma}(q)\, 
E^{\rho\sigma}(p_3,p_4),\label{calm}
\eea
where $q$ is the momentum transfered
\bea
q^\mu=p_1^\mu+p_2^\mu =p_3^\mu+p_4^\mu
\eea
and with $A^{\mu\nu|\rho\sigma}(q)$ solving the recursion equation
\bea
A^{\mu\nu|\rho\sigma}(q)=A_0^{\mu\nu|\rho\sigma}+{K^{\mu\nu}}_{\alpha\beta} A^{\alpha\beta|\rho\sigma}(q).\label{rec}
\eea
Lastly, in (\ref{rec}), $A_0$ is the contribution of the tree level amplitude derived from the quartic interaction vertex in (\ref{eq:Sexpansion}), 
\bea
A_0^{\mu\nu|\rho\sigma}=\frac{-1}{4V_0}\bigg(\eta^{\mu\rho}\eta^{\nu\sigma}+\eta^{\mu\sigma}\eta^{\nu\rho}-\eta^{\mu\nu}\eta^{\rho\sigma}\bigg)\equiv \frac{-1}{2 V_0} \Pi^{\mu\nu|\rho\sigma} \label{a0}
\eea
and $K$ is the so-called kernel
\begin{equation}
K^{\mu\nu}_{\quad \rho\sigma}= \frac{-iN}{4V_0}\Pi^{\mu\nu |\alpha\beta} \mathcal{I}_{\alpha\beta|\rho\sigma}\label{eqn:PV3kernel}
\end{equation}
with 
\begin{equation}
\mathcal I_{\alpha\beta|\rho\sigma} = -\int \frac{d^4p}{(2\pi)^4}\frac{ E^\mu_{\alpha\beta}(p+q,-p)E^\mu_{\rho\sigma}(p,-p-q)}{(p^2-m^2)((p+q)^2-\mu)} ,\label{iint}
\end{equation}
where the $E_{\alpha\beta} E_{\rho\sigma}$ factors account for the momentum dependence of the quartic vertices in the large $N$ limit{,  and where $ \Pi^{\mu\nu |\alpha\beta} $ is a simple rescaling of of the interaction vertex (\ref{a0}). }
The integral  in (\ref{iint}) is divergent by power counting. In \cite{Carone:2016tup} this divergence was regularized by working in $D=4-\epsilon$, and keeping $\epsilon$ small but finite.

\begin{figure}[h!]
\begin{center}
\includegraphics[scale=0.6]{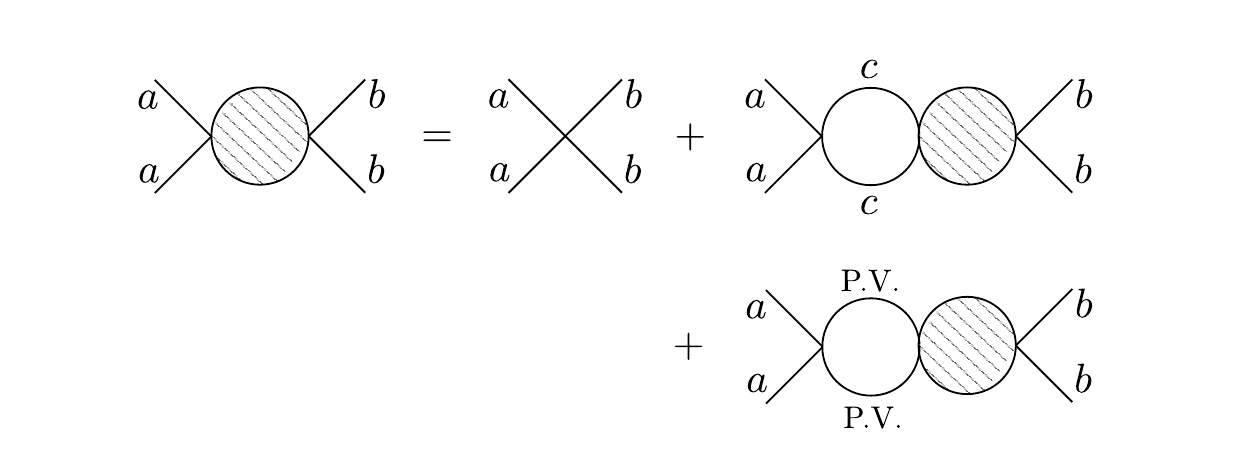}\caption{Two-into-two scalar scattering recursion relation, with Pauli-Villars fields. {The letters $a,b,c$ and P.V. indicate the species and type of the scalar fields. The sum is over all $N$-many intermediate $c$ scalar fields and all $N_{PV}$-many sets of Pauli-Villars fields.}} \label{recursive2}
\end{center}
\end{figure}

Instead, here we will use the Pauli-Villars fields to regularize \eqref{iint} as well as other one-loop divergences, such as the integrals that are canceled by the counterterms $c_1$ and $c_2$. As we will see in the next sections, it turns out that the minimum number of sets of Pauli-Villars fields $N_{PV}$ is three.  Each set of Pauli-Viallars fields will include $N$ fields, with the same mass and weights. We label these masses $M_1, M_2, \dots M_{N_{PV}}$. The weights are accounted by a set of coefficients $\alpha_1, \alpha_2\dots ,\alpha_{N_{PV}}$ which accompany each loop of a particular set of Pauli-Villars fields. In particular, the effect of adding the Pauli-Villars fields to the action (\ref{eq:S}) is to modify the kernel and $\mathcal I$  as in Figure \ref{recursive2}

\bea
\mathcal I_{\alpha\beta|\rho\sigma}(q) = -\int \frac{d^4p}{(2\pi)^4} \bigg[
\frac{ E_{\alpha\beta}(p+q,-p)E_{\rho\sigma}(p,-p-q)}{(p^2-m^2)((p+q)^2-m^2)} 
+ \sum_{i=1}^{N_{PV}} \frac{ \alpha_i E^{i}_{\alpha\beta}(p+q,-p)      
E^{i}_{\rho\sigma}(p,-p-q)}{(p^2-M^2_i)((p+q)^2-M^2_i)}\bigg]
,\nonumber\\ \label{eqn:I_general}
\eea
where
\bea
E^{i}{}^{\mu\nu}(p_1,p_2)\equiv-(p_1^\mu p_2^\nu+p_1^\nu p_2^\mu)+\eta^{\mu\nu}(p_1\cdot p_2+M^2_i).
\eea
{Notice the appearance of the weights $\alpha_i$ in (\ref{eqn:I_general}) in front of the Pauli-Villars contribution.}
\subsection{Zero momentum-transfer ($q=0$)}

To determine the masses and weights of the Pauli-Villars fields we first consider the zero momentum-transfer limit, $q^\mu=0$.  The reason is twofold: on the one hand, the regularization of the loop integrals is easily tractable, and on the other hand, we will use the results we derive for $q=0$ later in the next subsection to fine-tune $V_0$ such that the theory has a composite massless graviton.  

Let us then consider $N_{PV}$ sets of Pauli-Villars fields contributing to the scattering amplitude, at zero momentum transfer. We have
\begin{equation}
\mathcal I_{\alpha\beta|\rho\sigma}(q=0) = -\int \frac{d^4p}{(2\pi)^4}\bigg[ \frac{ E^\mu_{\alpha\beta}(p,-p)E^\mu_{\rho\sigma}(p,-p)}{(p^2-m^2)^2} + \sum_{i=1}^{N_{PV}} \frac{ \alpha_i E^{M^2_i}_{\alpha\beta}(p,-p)E^{M^2_i}_{\rho\sigma}(p,-p)}{(p^2-M^2_i)^2} \bigg]\label{key}
\end{equation}
Under the integral sign, Lorentz symmetry dictates that
\bea
\int d^4 p \, E_{\alpha\beta}(p,-p)E_{\rho\sigma}(p,-p) f(p^2)&=& \int d^4 p \bigg[
\frac{(p^2)^2}{6} (\eta_{\alpha\beta}\eta_{\rho\sigma}+\eta_{\alpha\sigma}\eta_{\beta\rho}+\eta_{\alpha\rho}\eta_{\beta\sigma})\nonumber\\
&&-m^2 \, p^2 \eta_{\alpha\beta}\eta_{\rho\sigma} + m^4\, \eta_{\alpha\beta}\eta_{\rho\sigma}\bigg] f(p^2)
\eea
for any Lorentz scalar function $f(p^2)$, and correspondingly,  {for each $i$,}
\bea
E^{M^2_i}_{\alpha\beta}(p,-p)E^{M^2_i}_{\rho\sigma}(p,-p) \times f(p^2)&=& \int d^4 p \bigg[ \frac{(p^2)^2}{6} (\eta_{\alpha\beta}\eta_{\rho\sigma}+\eta_{\alpha\sigma}\eta_{\beta\rho}+\eta_{\alpha\rho}\eta_{\beta\sigma})\nonumber\\
&&-M^2_i\ p^2 \eta_{\alpha\beta}\eta_{\rho\sigma} + M^4_i\ \eta_{\alpha\beta}\eta_{\rho\sigma}\bigg] f(p^2).
\eea
Substituting into \eqref{key} we have
\begin{align}
\mathcal I_{\alpha\beta|\rho\sigma}(q{=}0)\,=-\int \frac{d^4p}{(2\pi)^4} &\left[ \frac{(p^2)^2}{6} (\eta_{\alpha\beta}\eta_{\rho\sigma}+\eta_{\alpha\sigma}\eta_{\beta\rho}+\eta_{\alpha\rho}\eta_{\beta\sigma}) \left( \frac{1}{(p^2-m^2)^2} + \sum_{i=1}^{N_{PV}} \frac{\alpha_i}{(p^2-M^2_i)^2} \right)\nonumber \right.\\
& \!\!\!\!\!\!\!\!\!\!\!\!\!\!\!\!\!\!\!\!\!\!\!\!\!\!\!\!\!\!\!\!\!\!\!\!\!\!\!\!\!\!\!\!\!\!\!\!\!\!\!\!\!\!\!\!\left.-p^2\eta_{\alpha\beta}\eta_{\rho\sigma} \left( \frac{m^2}{(p^2-m^2)^2} + \sum_{i=1}^{N_{PV}} \frac{\alpha_i M^2_i}{(p^2-M^2_i)^2} \right) +\eta_{\alpha\beta}\eta_{\rho\sigma} \left( \frac{m^4}{(p^2-m^2)^2} + \sum_{i=1}^{N_{PV}} \frac{\alpha_i M^4_i}{(p^2-M^2_i)^2} \right)\right] \label{eqn:I}.
\end{align}
Counting the degree of divergence of the first term of \eqref{eqn:I}, we derive
\begin{equation}
{\rm{Div}}\int \frac{d^4p}{(2\pi)^4}\frac{(p^2)^2}{6} \ \left[ \frac{1}{(p^2-m^2)^2} + \sum_{i=1}^{N_{PV}} \frac{\alpha_i}{(p^2-M^2_i)^2} \right] \sim  \int_0^\infty dp \frac{p^7}{p^{2N_{PV}+4}}.\label{counting}
\end{equation}

To ensure the convergence of this integral, the minimum number of Pauli-Villars fields we need is $ N_{PV}=3 $. We also note that the Pauli-Villars fields will need to regularize the loops that lead to the counterterms $c_1$ and $c_2$, which we separately evaluate in Section \ref{c12}. The weights $\alpha_i$ for the case $ N_{PV}=3 $ are determined from
\bea
&1+\sum_{i=1}^{3} \alpha_i=0\nonumber\\
&m^2 +\sum_{i=1}^{3} \alpha_i M^2_i=0\nonumber\\
&m^4 + \sum_{i=1}^3 \alpha_i M^4_i=0.\label{conditionalpha}
\eea
These conditions arise from requiring that the power-counting divergent terms in \eqref{counting} are set to zero.\footnote{ Such equations are typical for Pauli-Villars regularized theories. For example, the first condition is typically written as $\sum c_i =0$, where $c_i$ are the statistical weights of all the fields (physical and regulators), and arises in regularizing the self-energy.}
Solving \eqref{conditionalpha} yields
\bea
&&\alpha_1=-\frac{(M^2_2-m^2)(M^2_3-m^2)}{(M^2_1-M^2_2)(M^2_1-M^2_3)}\nonumber\\
&&\alpha_2=-\frac{(M^2_1-m^2)(M^2_3-m^2)}{(M^2_2-M^2_1)(M^2_2-M^2_3)}\nonumber\\
&&\alpha_3=-\frac{(M^2_1-m^2)(M^2_2-m^2)}{(M^2_3-M^2_1)(M^2_3-M^2_2)}.
\label{eqn:fixed}
\eea
To simplify our calculations, we will further find useful to take the limit when the Pauli-Villars masses are all equal:
\bea M^2_{1,2,3}\to M^2 ,
\eea in which case the three sets  of terms in brackets in  (\ref{eqn:I}) become
\begin{align}
\frac{1}{(p^2-\mu)^2} + \sum_{i=1}^k \frac{\alpha_i}{(p^2-M^2_i)^2}  &= \dfrac{-(M^2-m^2)^3 (4p^2-M^2-3m^2) }{(p^2-m^2)^2(p^2-M^2)^{4}}\\
\frac{m^2}{(p^2-\mu)^2} + \sum_{i=1}^k \frac{\alpha_i M^2_i}{(p^2-M^2_i)^2} & =\dfrac{-(M^2-m^2)^3(3(p^2)^2-2p^2 m^2 -M^2 m^2) }{(p^2-m^2)^2(p^2-M^2)^{4}} \\
\frac{m^4}{(p^2-m^2)^2} + \sum_{i=1}^k \frac{\alpha_i M^4_i}{(p^2-M^2_i)^2} &=  \dfrac{-(M^2-m^2)^3 p^2 (2(p^2)^2+p^2(M^2-m^2)-2M^2 m^2) }{(p^2-m^2)^2(p^2-M^2)^{4}}
\end{align}
It is now straightforward to combine the denominators with just one Feynman parameter and compute the momentum integral using
\begin{equation}
\int \dfrac{d^4p}{(2\pi)^4} \dfrac{(p^2)^{n_1}}{(p^2-\Delta)^{n_2}}= \dfrac{i 2\pi^2}{(2\pi)^4} (-1)^{n_1+n_2} \int_0^\infty dp \dfrac{p^3 p^{2n_1}}{(p^2+\Delta)^{n_2}}, \label{eqn:euclidean_integration}
\end{equation}
where the factor of $i$ is picked up after the Wick rotation, and the factors of $-1$ arise due to the choice of mostly minus Minkowski metric (so $p^2$ in Minkowski signature becomes $-p^2$ after the Wick rotation).
Subsequent integration over the only Feynman parameter in the game yields
\begin{align}\label{qzero}
\mathcal I_{\alpha\beta|\rho\sigma}(q{=}0) \,=
&\dfrac{i}{64\pi^2 } \left(M^4-4M^2 m^2+2m^4\ln(M^2/m^2)+3m^4\right)
(\eta_{\alpha\sigma}\eta_{\beta\rho}+ \eta_{\alpha\rho}\eta_{\beta\sigma}-\eta_{\alpha\beta}\eta_{\rho\sigma}).
\end{align}
We would like to point out that this is an exact result, derived without taking any limit on the Pauli-Villars  mass $M$. 
If we take the limit of the Pauli-Villars mass much larger than the mass of the physical scalars $M^2\gg m^2$, we get 
\bea
\mathcal I _{\alpha\beta|\rho\sigma}(q{=}0)\, = M^4\bigg[ \frac{i}{64\pi^2}(\eta_{\alpha\sigma}\eta_{\beta\rho}+ \eta_{\alpha\rho}\eta_{\beta\sigma}-\eta_{\alpha\beta}\eta_{\rho\sigma})  + \mathcal{O}(M^{-2})\bigg].
\eea


\subsection{Small momentum-transfer ($q\neq 0$) and the graviton pole  }

In the general case, we should compute the same kernel as equation \eqref{eqn:PV3kernel}, but this time with {the transferred momentum $q\neq 0$ appearing in the  $E_{\alpha \beta }$ factors}
\begin{equation}
\mathcal I_{\alpha\beta|\rho\sigma}(q) = -\int \frac{d^4p}{(2\pi)^4} \frac{ E_{\alpha\beta}(p+q,-p)E_{\rho\sigma}(p,-p-q)}{(p^2-m^2)((p+q)^2-m^2)} + \sum_{i=1}^{3} \frac{ \alpha_i E^{M^2_i}_{\alpha\beta}(p+q,-p)E^{M^2_i}_{\rho\sigma}(p,-p-q)}{(p^2-M^2_i)((p+q)^2-M^2_i)}. \label{eqn:I_general}
\end{equation}
We use a minimal set of Pauli-Villars fields ($N_{PV}=3$), with weights $\alpha_i$ which were determined previously in \eqref{eqn:fixed}. 

We first combine the denominators of  \eqref{eqn:I_general} using a Feynman parameter, so the integral above looks like equation \eqref{eqn:I}:
\begin{align}
	\dfrac{1}{(p^2-m^2)((p+q)^2-m^2)}&=\int_0^1 dx\ \dfrac{1}{\bqty{(1-x)(p^2-m^2)+x((p+q)^2-m^2)}^2}=\int_0^1 dx\ \dfrac{1}{\bqty{k^2-\Theta}^2} \label{eqn:general_xfeyn} \\
	\dfrac{1}{(p^2-M^2_i)((p+q)^2-M^2_i)}&=\int_0^1 dx\ \dfrac{1}{\bqty{(1-x)(p^2-M^2_i)+x((p+q)^2-M^2_i)}^2}=\int_0^1 dx\ \dfrac{1}{\bqty{k^2-\Theta_i}^2}\label{eqn:general_xfeyn_PV}\, ,
\end{align}
where
\begin{equation}
	k^\mu=p^\mu + x q^\mu, \qquad \Theta = \mu - (1-x)x q^2 \qand \Theta_i = M^2_i - (1-x)x q^2\, .
\end{equation}
After shifting the integration to $ k^\mu $, we then have
\begin{align}
\mathcal I_{\alpha\beta|\rho\sigma}(q) = -\int_0^1 dx\ &\frac{d^4k}{(2\pi)^4}  \left\lbrace \frac{ E_{\alpha\beta}(k+(1-x)q,xq-k) E_{\rho\sigma}(k-xq,(x-1)q-k)}{\bqty{k^2-\Theta}^2} \right. \nonumber \\
& \left. + \sum_{i=1}^{3} \frac{ \alpha_i E^{M^2_i}_{\alpha\beta}(k+(1-x)q,xq-k) E^{M^2_i}_{\rho\sigma}(k-xq,(x-1)q-k)}{\bqty{k^2-\Theta_i}^2} \right\rbrace \label{eqn:I_xfeyn}\, .
\end{align}
Expanding the numerators and re-arranging in powers of $ k^2 $, we can cast this equation into a form analogous to \eqref{eqn:I}:
\begin{align}
\mathcal I_{\alpha\beta|\rho\sigma}=-\int_0^1 dx\ \frac{d^4k}{(2\pi)^4} & 
\bigg\{(k^2)^2 F_{\alpha\beta;\rho\sigma}(q,x) \left[ \frac{1}{(k^2-\Theta)^2} + \sum_{i=1}^3 \frac{\alpha_i}{(k^2-\Theta_i)^2} \right] \nonumber\\&+ k^2 G_{\alpha\beta;\rho\sigma}(q,x) \left[ \frac{m^2}{(k^2-\Theta)^2} + \sum_{i=1}^3 \frac{\alpha_i M^2_i}{(k^2-\Theta_i)^2} \right]\nonumber\\
& + H_{\alpha\beta;\rho\sigma}(q,x) \left[ \frac{m^4}{(k^2-\Theta)^2} + \sum_{i=1}^3 \frac{\alpha_i M^4_i}{(k^2-\Theta_i)^2} \right]\bigg\} \label{eqn:I_general_bracket_form}\, ,
\end{align}
where $ F_{\alpha\beta;\rho\sigma}(q,x), G_{\alpha\beta;\rho\sigma}(q,x), H_{\alpha\beta;\rho\sigma}(q,x) $ are {some} functions of $ q^\mu $ and $ x $, but strictly not of $ k^\mu $.  
After performing the loop momentum integrals (at the cost of introducing one more Feynman parameter), and in the limit  $q^2\ll m^2$, we finally  arrive at
\begin{align}
\mathcal I_{\alpha\beta|\rho\sigma}(q)& \simeq -\dfrac{iM^4}{64\pi^2}(\eta_{\alpha\sigma}\eta_{\beta\rho}+\eta_{\alpha\rho}\eta_{\beta\sigma}-\eta_{\alpha\beta}\eta_{\rho\sigma}) 
+ \dfrac{iM^2}{192\pi^2} \big[(12 m^2-q^2) ( \eta_{\alpha\sigma}\eta_{\beta\rho} +\eta_{\alpha\rho}\eta_{\beta\sigma} -\eta_{\alpha\beta}\eta_{\rho\sigma} ) 
\nonumber \\
&  + q_\alpha q_\sigma \eta_{\beta\rho} + q_\alpha q_\rho \eta_{\beta\sigma} + q_\beta q_\rho \eta_{\alpha\sigma} + q_\beta q_\sigma \eta_{\alpha\rho} -2q_\alpha q_\beta \eta_{\rho\sigma} - 2q_\rho q_\sigma \eta_{\alpha\beta} \big]\, .
\end{align}
We can now derive the kernel for the recursive iteration of the scattering amplitude
\begin{equation}
A^{\mu\nu|\alpha\beta} = A_0^{\mu\nu|\alpha\beta} + K^{\mu\nu}_{\quad \rho\sigma} A^{\rho\sigma|\alpha\beta}\label{rec11},
\end{equation} where $A_0$ was defined previously in \eqref{a0}, and the kernel $K$ equals
\begin{equation}\label{k3}
K^{\mu\nu}_{\quad \rho\sigma}(q)= \frac{-iN}{4V_0}\Pi^{\mu\nu | \alpha\beta} \mathcal{I}_{\alpha\beta|\rho\sigma} \simeq \frac{NM^2}{4V_0} \bqty{\dfrac{M^2}{64\pi^2} - \dfrac{ q^2+12 m^2 }{192\pi^2}}(\delta^\mu_\sigma \delta^\nu_\rho + \delta^\mu_\rho \delta^\nu_\sigma).
\end{equation}

Given the tensor structure of the kernel \eqref{k3}, we solve the recursion relation \eqref{rec11} by
matching the tensor structure of the tree level amplitude and solving for the overall function:
\begin{equation}
A^{\mu\nu|\alpha\beta} = a(q^2)(\eta^{\mu\alpha} \eta^{\nu\beta} + \eta^{\mu\beta} \eta^{\nu\alpha} - \eta^{\mu\nu} \eta^{\alpha\beta}) \, .\label{ansatz}
\end{equation}
If we fine-tune the value of $V_0$
\bea 
V_0\simeq  \dfrac{M^2 N (M^2 - 4 m^2)}{128\pi^2} \, \label{eqn:3Pv0},
\eea
then we identify a massless pole in the scattering amplitude,  with $a(q^2)$ given by
\begin{equation}
	a(q^2)\simeq -\dfrac{96\pi^2}{M^2 N}\frac{1}{q^2}  \label{eqn:3Pvab}\, ,
\end{equation}
where we recall that $M$ is the mass of the Pauli-Villars fields.

We now note that the scattering amplitude in \eqref{calm}, at low $q^2$,  with $A^{\mu\nu|\alpha\beta}$ given by \eqref{ansatz} and  \eqref{eqn:3Pvab}  provides the evidence for a composite massless spin 2 particle (the graviton) in the spectrum of our theory: it is of the same form as the two-into-two scalar tree-level scattering amplitude mediated by a graviton exchange in general relativity, in which case $A^{\mu\nu|\alpha\beta}$ corresponds to the graviton propagator multiplied by the gravitational coupling constant
\bea
A^{\mu\nu|\alpha\beta}_{\rm graviton}(q)=-\frac{1}{2M_{\Pl}^2}\dfrac{1}{q^2}(\eta^{\mu\alpha} \eta^{\nu\beta} + \eta^{\mu\beta} \eta^{\nu\alpha} - \eta^{\mu\nu} \eta^{\alpha\beta}) .
\eea
In terms of the Pauli-Villars scale $M^2$, the Planck mass is given by
\bea
M_{\Pl}\simeq \sqrt{\frac{M^2 N}{192 \pi^2}}\label{planckpv}.
\eea
We conclude that the presence of the composite graviton in this model is not an artefact of the regularization scheme chosen in the earlier work. It is however conditioned by a similar fine-tuning of the constant $V_0$.
The Planck mass has a value that is determined by the regulator, with \eqref{eq:mpscat} replaced by \eqref{planckpv}.

We can actually do a better job and find the exact (not just leading order in $M^2$) value for $V_0$ such that the spectrum of the theory contains the massless graviton. To do this we need the zero transfer-momentum expression for $\mathcal I_{\alpha\beta | \rho\sigma}$ given in \eqref{qzero}. First we note that 
\bea
-\frac{i N}{4V_0}\Pi^{\mu\nu| \rho\sigma}{\mathcal I}_{\rho\sigma | \alpha\beta }(q{=}0)=\dfrac{N}{2^8 \pi^2  V_0} \left(M^4-4M^2 m^2+3m^4+2m^4\ln(M^2/m^2)\right)(\delta^\mu_\alpha \delta^\nu_\beta+\delta^\mu_\beta \delta^\nu_\alpha) \, .\nonumber\\
\eea
Next, the recursion relation $A=A_0+K\cdot A$ can be manipulated by moving the $q=0$ terms in the kernel to the left hand side:
\bea\label{vzero}
&&A^{\mu\nu|\alpha\beta}\bigg(1-\dfrac{N}{128 \pi^2 V_0} \left(M^4-4M^2 m^2+3m^4+2m^4\ln(M^2/m^2)\right)\bigg)\nonumber\\&&
=A_0^{\mu\nu|\alpha\beta} -\frac{N M^2}{384 \pi^2 V_0} \bigg(q^2 +{\cal O}(q^4)\bigg) A^{\mu\nu|\alpha\beta}.
\label{higherderivpole}
\eea
If the left hand side  of \eqref{vzero} is zero, then we can immediately solve for the scattering amplitude and obtain a pole at $q^2=0$, with the correct tensor structure to identify it as the propagator of a massless spin 2 particle, the composite graviton. Therefore the existence of the massless composite spin two particle is conditioned by the fine tuning
\bea
V_0=\dfrac{N}{128 \pi^2 } \left(M^4-4M^2 m^2+2m^4\ln(M^2/m^2)+3 m^4\right)\equiv \frac{N M^4}{128\pi^2} {\cal V}_3\label{v03exact}.
\eea
If $V_0$ does not satisfy the fine-tuning condition  \eqref{v03exact}, then the amplitude takes the form
\bea
A^{\mu\nu|\alpha\beta}(q)\simeq -\bigg(\frac{1}{NM^2/(96 \pi^2)} \bigg)
\frac{\eta^{\mu\alpha} \eta^{\nu\beta} + \eta^{\mu\beta} \eta^{\nu\alpha} - \eta^{\mu\nu} \eta^{\alpha\beta}}{q^2  -3 M^2 {\cal V}_3 + 384 \pi^2V_0/ (NM^2)} \,.
\eea
For values of  $V_0$ such that $V_0<NM^4{\cal V}_3/(128 \pi^2 )$, summing up the diagrams in Figure \ref{recursive2}, leads to an effective exchange of massive spin 2  and spin 0 particles, with equal masses \cite{VanNieuwenhuizen:1973fi}. We infer the existence of the spin 0  particle, its mass and that it is a {\it ghost} from the tensor structure of the scattering amplitude $A^{\mu\nu|\alpha\beta}$. So, in this case, the model has a composite massive spin 2 particle, and a composite ghost scalar\footnote{A linearized theory involving a massive symmetric two-index tensor $h_{\mu\nu}$ propagates only the 5 degrees of freedom of the massive graviton if the mass term has the Fierz-Pauli structure $\mu^2[ (h_\mu^\mu)^2- h_{\mu\nu} h^{\mu\nu}]$. If the relative coefficient between $h^2$ and $h_{\mu\nu}h^{\mu\nu}$ is not -1, then the theory will contain a scalar ghost. For clarity, and referring to section 4
\cite{VanNieuwenhuizen:1973fi}, the propagator of a massive symmetric 2-index tensor is given in equation (50). $P^2$ is the propagator of a massive graviton (note that it has a different tensor structure than that of a massless graviton) and $S$ is the scalar propagator. When inserted between conserved sources, as in our problem, the $k$-dependence cancels out, and the $\omega$-tensors drop out while the $\theta$-tensors reduce to the Minkovski metric. For $a=1, b=1, c=1, d=-1$ and $\beta=-1$ (where $\beta$ is the relative coefficient between the two mass terms) the theory reduces to Fierz-Pauli and there is no contribution coming from $S$. However, if $\beta=1/2$, the scalar has the same mass as the spin 2 field, $\mu^2$, and the contribution from the scalar propagator together with the massive spin 2 Fierz-Pauli proagator combine to give the tensor structure of the massless graviton, as we've seen in our own scattering amplitude $A^{\mu\nu|\alpha\beta}$.}.
This observation was also made in \cite{Carone:2019xot}.
Lastly, if $V_0>NM^4{\cal V}_3/(128 \pi^2) $ there are no composite particle poles at low momentum transfer.

\subsection{A non-minimal set of Pauli-Villars fields: $N_{PV}=4$}\label{4pv}

In this subsection we investigate the consequences of using a non-minimal set of Pauli-Villars fields. Let's consider the case of four sets of Pauli-Villars fields whose statistical weights $ \alpha_i $ satisfy
the regularization conditions encountered earlier 
\bea
1+\sum_{i=1}^4\alpha_i=0, \qquad m^2+\sum_{i=1}^4\alpha_iM^2_i=0,\qquad m^4+\sum_{i=1}^4\alpha_iM^4_i=0,
\eea
together
with one additional constraint
\bea
m^6+\sum_{i=1}^4\alpha_iM^6_i=0.
\eea
With more Pauli-Villars fields we are now rendering the loop integrals convergent in a faster way,  rather than ensuring their mere convergence. 
This leads to the following weights for our non-minimal set of Pauli-Villars fields
\bea
&&\alpha_1=\frac{(M^2_2-m^2(M^2_3-m^2)(M^2_4-m^2)}{(M^2_1-M^2_2)(M^2_1-M^2_3)(M^2_1-M^2_4)\nonumber }\\
&&\alpha_2=\frac{(M^2_1-m^2)(M^2_3-m^2)(M^2_4-m^2)}{(M^2_2-M^2_1)(M^2_2-M^2_3)(M^2_2-M^2_4)} \nonumber \\
&&\alpha_3=\frac{(M^2_1-m^2)(M^2_2-m^2)(M^2_4-m^2)}{(M^2_3-M^2_1)(M^2_3-M^2_2)(M^2_3-M^2_4)}\nonumber \\
&&\alpha_4=\frac{(M^2_1-m^2)(M^2_2-m^2)(M^2_3-m^2)}{(M^2_4-M^2_1)(M^2_4-M^2_2)(M^2_4-M^2_3)}\,.
\eea

We will follow the previous analysis to compute the kernel
\bea
K^{\mu\nu}_{\quad \rho\sigma}(q)=\frac{-iN}{4V_0}\Pi^{\mu\nu |\alpha\beta} \mathcal{I}_{\alpha\beta|\rho\sigma}(q)\, ,\nonumber
\eea
where now we  have to include the contribution of four Pauli-Villars fields running in the loop
\begin{equation}
\mathcal I_{\alpha\beta|\rho\sigma} (q)= -\int \frac{d^4p}{(2\pi)^4} \frac{ E_{\alpha\beta}(p+q,-p)E_{\rho\sigma}(p,-p-q)}{(p^2-m^2)((p+q)^2-m^2)} + \sum_{i=1}^{4} \frac{ \alpha_i E^{M^2_i}_{\alpha\beta}(p+q,-p)E^{M^2_i}_{\rho\sigma}(p,-p-q)}{(p^2-M^2_i)((p+q)^2-M^2_i)}\, .\label{eqn:4PV_I}
\end{equation}
In the zero-momentum transfer  limit ($q^\mu=0$) we find the exact result
\begin{equation}
\mathcal I_{\alpha\beta|\rho\sigma}(q{=}0) \,=
\dfrac{i}{192\pi^2} \pqty{M^4-6M^2 m^2 +3m^4 +6m^4 \ln(M^2/m^2)+\frac{2m^6}{M^2}}
(\eta_{\alpha\sigma}\eta_{\beta\rho}+ \eta_{\alpha\rho}\eta_{\beta\sigma}-\eta_{\alpha\beta}\eta_{\rho\sigma})\, .
\end{equation}

The kernel, in the limit of small $q^2$, is given by:
\begin{equation}
K^{\mu\nu}_{\quad \rho\sigma}(q)=\frac{-iN}{4V_0}\Pi^{\mu\nu | \alpha\beta} \mathcal{I}_{\alpha\beta|\rho\sigma} \simeq \pqty{\frac{NM^2}{4V_0}}\bqty{ \frac{M^2}{192\pi^2}-\frac{q^2+9m^2} {288\pi^2}} \pqty{\delta^\mu_\sigma \delta^\nu_\rho + \delta^\mu_\rho \delta^\nu_\sigma}, \label{kernel4pv}
\end{equation}
where we took the convenient limit of equal  Pauli-Villars masses  $M^2_i \longrightarrow M^2$.

We solve for the scattering amplitude $A^{\mu\nu|\alpha\beta}(q)=A_0^{
\mu\nu|\alpha\beta}+K^{\mu\nu}_{\quad \rho\sigma}(q)A^{\rho\sigma|\alpha\beta}(q)$. Rearranging by moving the $q=0$ terms in $K\cdot A$ to the left hand side we find
\bea
&&A^{\mu\nu|\alpha\beta}\bigg(1-\frac{N}{384 \pi^2 V_0}\bqty{M^4-6M^2 m^2 +6m^4 \ln (\frac{M^2}{m^2})+3m^4+\frac{2m^6}{M^2}} \bigg)\nonumber\\
&&= A_0^{\mu\nu|\alpha\beta} - \frac{N M^2}{576 \pi^2 V_0}(q^2 + {\cal O}(q^4))A^{\mu\nu|\alpha\beta}.\label{a4pv0}
\eea
We infer from here that 

(i) if we fine-tune $V_0$ such that
\begin{equation}
V_0=\frac{N}{384\pi^2}\bqty{M^4-6M^2 m^2 +6m^4 \ln (\frac{M^2}{m^2})+3m^4+\frac{2m^6}{M^2}} \equiv\frac{NM^4}{384\pi^2} {\cal V}_4 \label{4Pvv0},
\end{equation}
then, from \eqref{a4pv0}) we see that the scattering amplitude has a massless pole
\begin{equation}
 A^{\mu\nu|\alpha\beta} \simeq -\frac{144\pi^2}{M^2 N}\frac{1}{q^2} (\eta^{\mu\alpha}\eta^{\nu\beta}+
\eta^{\mu\beta}\eta^{\nu\alpha} - \eta^{\mu\nu}\eta^{\alpha\beta})\, .\label{apv4}
\end{equation}
The tensor structure of (\eqref{apv4}) identifies the massless particle as a spin-two particle. Hence the spectrum retains the composite massless graviton. However the values of the fine-tuned $V_0$ and the Planck mass, with $M_{\Pl}=\sqrt{M^2 N/(288 \pi^2)}$, are sensitive to the number of regulator fields.

(ii) if we do not set $V_0$ equal to its fine-tuned value \eqref{4Pvv0}, then the scattering amplitude is given by:
\bea
A^{\mu\nu|\alpha\beta}\simeq
\bigg(\frac{-1}{NM^2/(144 \pi^2)} \bigg)
\frac{\eta^{\mu\alpha} \eta^{\nu\beta} + \eta^{\mu\beta} \eta^{\nu\alpha} - \eta^{\mu\nu} \eta^{\alpha\beta}}{q^2  -3/2 M^2 {\cal V}_4+ 576 \pi^2V_0/ (NM^2)} \,.\nonumber\\
\eea
As seen before, without the fine-tuning condition, the scattering amplitude has a pole corresponding to the exchange of massive spin 2 and ghost spin 0 particles provided that $V_0<NM^4{\cal V}_4/(384\pi^2) $. Otherwise, there are no composite particles.

\section{Conclusion}

We have shown that the existence of a massless composite graviton in the non-metric, non-polynomial scalar action
\eqref{eq:S}, introduced earlier by Carone, Erlich and Vaman in \cite{Carone:2016tup} is not dependent on the choice of regulator, as long as the regulator respects general covariance. In this work we have replaced the use of dimensional regularization, employed in previous works \cite{Carone:2016tup,Carone:2017mdw,Carone:2018ynf},  with Pauli-Villars fields, using the functional determinant prescription of \cite{Diaz:1989nx}. We worked with a minimal set of Pauli-Villars fields, and we considered the implications of a non-minimal set of Pauli-Villars fields. In each case we found that there exists a fine-tuning condition which results in a massless graviton pole in the low momentum transfer scattering amplitude of two-into-two scalars. In each case, the Planck constant can be extracted, and it is dependent on the regulator masses.

 If, on the other hand, the fine-tuning choice (\ref{v03exact}) is not made, then from the form of the scattering amplitude we conclude that the theory has a composite massive spin 2 state and a spin 0 ghost of equal mass \cite{VanNieuwenhuizen:1973fi}. Interestingly, in general relativity this situation arises in perturbation theory when the cosmological constant is non-zero and one truncates to second order in fluctuations around flat space \cite{Gabadadze:2003jq}.\footnote{
As  in \cite{Gabadadze:2003jq}, there is a redefinition of the metric fluctuations which gets rid of the term linearized in fluctuations and casts the linearized action into the 
linearized action of a massive symmetric two-index tensor with a relative coeffcient between the two mass terms $\beta =-1/2$ \cite{VanNieuwenhuizen:1973fi}, as discussed previously in Footnote 4.}
Therefore, we see once more\footnote{In \cite{Carone:2016tup} a different argument based on a non-cancellation of tadpole diagrams was given to argue that the fine-tuning of $V_0$ is correlated to the vanishing of the cosmological constant.}  that the fine-tuning of the potential $V_0$ is the same as the fine-tuning which sets the cosmological constant to zero.

We would like to point out another aspect of our work. At no step in our calculations did we take the infinite mass limit for the Pauli-Villars fields. For example, the fine-tuning condition \eqref{v03exact} is an exact expression in the masses, at large $N$. Theories with finite-mass Pauli-Villars fields are Lee-Wick theories \cite{Lee:1970iw}. Such theories also arise in describing higher-derivative theories of  gravity. The composite gravity model analyzed in this paper does, in fact, lead to higher-derivative interactions of the composite graviton. These are expected given that there are terms of order $q^4$ in the equation \eqref{higherderivpole}, which imply that the composite graviton propagator pole is of the form $1/((q^2)(q^2+\Delta))$.

In the series of works  \cite{Carone:2016tup,Carone:2017mdw,Carone:2018ynf} the analysis which led to the identification of  a composite graviton was performed in the simplifying large $N$ limit (where $N$ is the number of physical scalar fields). Therefore we think that it would be valuable to investigate the consequences of moving away from this limit for the existence of the composite graviton.

\acknowledgements

The authors are indebted to Chris Carone and Joshua Erlich for many enlightening conversations.

\begin{appendix}
\section{Clock-and-ruler fields, $c_1$ and $c_2$ counterterms}\label{c12}

\begin{figure}[h!]
\includegraphics[scale=0.6]{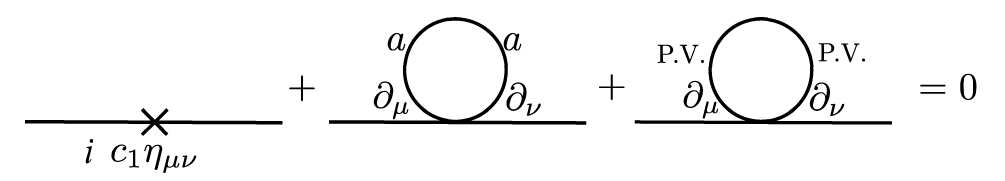}\\
\includegraphics[scale=0.6]{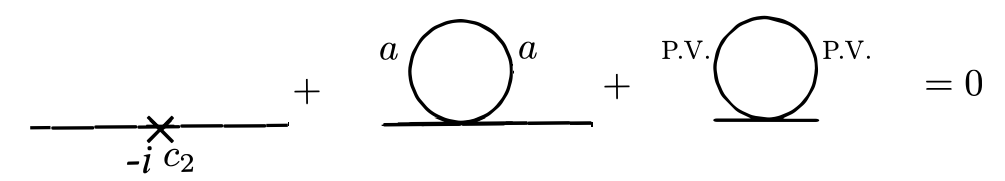}\caption{Counterterms}\label{c12fig}
\end{figure}

The recursion relation for the scattering amplitude and of the kernel in the previous section hinge on being able to resum, or eliminate by the addition to the original Lagrangian of counterterms, those loop diagrams which would renormalize the mass and the wave function. See Figure \ref{c12fig}.
The goal of this section is to compute $c_1$ and $c_2$ using our set of Pauli-Villars fields as regulators, where the choices for the statistical weights $\alpha$ and masses have already been  made. 
\subsection{ Three Pauli-Villars fields}
For $c_1$ we demand that the following vacuum expectation value vanishes
\bea
\langle -c_1\eta_{\mu\nu}+\partial_\mu\phi^a\partial_\nu \phi^a+\sum_{i=1}^3 \partial_\mu\Phi_{i, PV}^a\partial_\nu \Phi_{i,PV}^a\rangle=0\, ,
\eea
which implies that
\bea
- c_1\eta_{\mu\nu} + (-i)(i)(i)
N\frac{\eta_{\mu\nu} }{4} \int \dfrac{d^4k}{(2\pi)^4} k^2 \pqty{\dfrac{1}{k^2-m^2}+ \sum_{i=1}^{3} \dfrac{\alpha_i}{k^2-M^2_i}} =0\, .\label{eqn:k^2micky}
\eea
Therefore, $c_1$ evaluates to
\bea
c_1=-\dfrac{N}{128\pi^2} \bqty{M^4-4m^2 M^2 +2m^4 \ln(\frac{M^2}{m^2})+ 3m^4} ,\label{c1det}
\eea
in the limit of equal Pauli-Villars masses.
For $c_2$, we impose
\bea
\langle c_2 - \frac 12(m^2 \phi^a \phi^a + \sum_{i=1}^{3} M^2_i \Phi^a_{i,PV} \Phi^a_{i,PV})\rangle =0\, .
\eea
At one loop order, this means that
\begin{equation}
c_2-
\frac{iN}2\int \dfrac{d^4k}{(2\pi)^4} \pqty{\dfrac{m^2}{k^2-m^2}+ \sum_{i=1}^{3} \dfrac{\alpha_i M^2 _i}{k^2-M^2_i}}=0 \label{eqn:m^2micky}\, ,
\end{equation}
which yields
\bea
c_2=\dfrac{N}{64\pi^2} \bqty{M^4-4m^2 M^2+2m^4\ln(\frac{M^2}{m^2}) + 3m^4} \, ,
\eea
in the same limit of equal Pauli-Villars masses.

Consider now the  case when the spectrum of the thery contains a massless graviton, corresponding to fine-tuning $V_0$ to the value 
\eqref{v03exact}.  Then, the 
the gauge-fixing condition of the clock and ruler fields \eqref{eq:staticgauge},  where  $c_1$ is computed above in \eqref{c1det},  sets these fields to zero. This is a feature that was already encountered when the theory was regularized by working in $D=4-\epsilon$, with $\epsilon$ taken to be finite (small and negative, to be precise).

\subsection{Four Pauli-Villars fields}

Working out $c_1$ and $c_2$ with four Pauli-Villars fields, with masses and weights as in subsection \ref{4pv}, we obtain
\bea
c_1 = -\frac{N}{384\pi^2}\bqty{M^4-6M^2 m^2+6m^4\ln(\frac{M^2}{m^2})+3m^4+\frac{2m^6}{M^2}}
\eea
and
\bea
c_2=\frac{i}{192\pi^2}\bqty{M^4-6M^2 m^2+6m^4\ln(\frac{M^2}{m^2})+3m^4+\frac{2m^6}{M^2}}\, .
\eea
Once more, we see that $V_0-c_1=0$ if $V_0$ is fine-tuned as in \eqref{4Pvv0}, which is the case when there is a composite massless graviton in the spectrum.

\end{appendix}


\end{document}